\documentclass[12pt,epsf]{article}

\usepackage{amsmath,amssymb,graphicx}

\newcommand{\beq}{\begin{eqnarray}}% can be used as {equation} or  {eqnarray}
\newcommand{\eeq}{\end{eqnarray}}

\title{
{\huge % \bf One two-site  CHM with Composite Fermions} \vspace*{0.8cm}
\bf  Vector-like Fermions  in  a  Minimal Composite Higgs Model } \vspace*{0.8cm}
 }

\author{Haiying Cai\\
 \\ \normalsize\emph{Department of Physics, Peking University, Beijing 100871, China}  \\}

\date{}   %{\today}
\begin{document}
\setcounter{page}{0} \maketitle
\thispagestyle{empty}
\vspace*{0.5cm} \maketitle  %\begin{center}

\begin{abstract}
We  consider  the scenario where the composite Higgs  arising as  a pNGB in a two-site model with a non-local term included.  Constraints from  pion scattering and  electroweak precision test are  considered.  We discuss  the effects of  composite resonances, in particular the one from  composite vector-like fermions,  on the  oblique parameters. It is  noticed that the  gluon fusion production of  Higgs boson  is  suppressed with respect to  the Standard Model  for  about  $6\%$  after  imposing  the unitarity and  electroweak bounds.
\end{abstract}

\thispagestyle{empty}

\newpage

\setcounter{page}{1}

\section{ Introduction}
With  the discovery of the $125~\mbox{GeV}$ Higgs boson at the LHC~\cite{Higgs},  one of the most  crucial task is to unveil the nature of  this scalar particle. Current measurement  of   Higgs couplings in various channels reports  certain deviation from the SM  expectation,  implicating that  new physics  may exist  beyond the TeV energy scale,  although there is no  direct  signature  that confirms the existence of  new particles.  One of  the theoretically plausible  frameworks  for the  BSM  new physics is  the traditional SUSY,  which aims to solve the hierarchy problem and propose mechanisms for  the electroweak symmetry breaking. In  this paper, we are interested to explore the composite Higgs scenario as an alternative option, since it will provide a  little  Higgs candidate  from an underlying  strong dynamic sector. There are many varieties of composite Higgs models, with the original one  realized in an extra dimensional scenario following the AdS/CFT correspondence~\cite{CHM}.  The Higgs  potential is  calculable in a 5D description  of  composite Higgs Model, but the particle spectrum in this framework is much more complicated and not easy  to make contact with the LHC measurement.  Therefore recent effort is more focused on the 4D  construction of  Composite Higgs  theory (CHM)~\cite{Anastasiou:2009rv, Panico,DeCurtis:2011yx}.  Since only the lowest lying  states are accessible  in  the future  LHC, it is adequate to formulate a  predictive description without resorting to an  UV  completing picture.

One effective approach to qualitatively describe a  strong dynamic theory is employing  the Callan-Coleman-Wess-Zumino  (CCWZ) formalism~\cite{CCWZ}, where the  full global symmetry is nonlinearly realized and the Lagrangian is constructed by  covariant objects transforming under a local symmetry group. The CCWZ prescription captures the common features in a generic CHM  e.g. modified Higgs couplings,  while on the other hand it hides the dynamic origin of the partial compositeness  and leaves  the masses of  vector bosons  to  be  less correlated.  Here we  are going to use  the deconstruction method proposed by~\cite{Panico, DeCurtis:2011yx}  to  parametrize  the scenario where  one composite Higgs  is realized as a pNGB  from a spontaneous breaking global symmetry. We are interested in exploring the simplest case, a two-site model with an enlarged global symmetry of  $SO(5)_1 \times SO(5)_2$.  In such a  description, only the first level of  composite resonances is available.  In contrast to the CCWZ formalism, the symmetry is realized in a linear way via the  deconstruction method, with the partial compositeness  being manifested as  the  result  of symmetry breaking in the composite sector. In particular non-local terms are possible to be introduced into this two-site model according to the symmetry principle.  The existence of non-local terms has crucial  impact on the unitarity of  $W_L W_L$ scattering and  could  change the sign of $S$ parameter under certain condition. Composite vector-like fermions  in  $SO(5)$ representations are necessarily to be incorporated so that the SM  fermions will gain the masses.  One important motivation for this paper is to explore the influence of  composite  fermions on  the electroweak precision test  and estimate their contribution to  the Higgs production.

The paper  is organized in the following  way.  We starts from a  review of the model set up in the two-site description in Section 2.  For the gauge bosons, the spectrum is calculated in the unitary gauge and we investigate the influence of  non-local term on the $\pi \pi \to \pi \pi$  scattering.  While for the fermion sector,  composite fermions in a basic  $SO(5)$ representation are included and we are going to explore their mixing with the  SM  fermions.  In Section 3,  the contribution to $S$ and $T$ parameter from the vector and fermion resonances are illustrated in detail, which is further compared with the  experimental data  by a numerical scanning of the parameter space in this two-site model.   Finally in Section 4,   we  estimate  the  reduced  Higgs production rate from the gluon fusion process  by imposing the EW constraints on the composite scale $f$.

\section{two-site model}
\begin{figure}[h]
\begin{center}
\includegraphics[angle=0,clip,width=7cm]{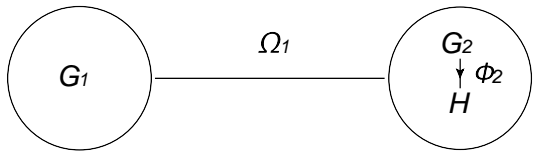}
\caption{{\small   moose diagram for a two-site model, where the global symmetry is $G_{1} \times G_{2} = SO(5)_{1} \times SO(5)_{2}$.
The two sites are connected by a link field $\Omega_1$ and  the $SO(5)_2$ symmetry
is spontaneously broken by a  local field $\Phi_2 = \Omega_2 \phi_0$ into a subgroup $H= SO(4)$.}} \label{twosite}
\end{center}
\end{figure}
Let us first  review the basic  model  set up.  The two-site model is the simplest  scenario  to describe a composite Higgs boson,  with one site imitating the  UV brane and the other site imitating the bulk in a warped extra dimension.  Since the symmetry breaking is generalized to be  $SO(5)_1 \times SO(5)_2/SO(4)$ as depicted in Fig.~\ref{twosite}, there are  in principle  two  sets of  non-linear sigma fields  in order to describe the  coset space.  One link field $\Omega_1$ mediating the interaction between the site-1 and the site-2 will break the $SO(5)_1 \times SO(5)_2$ into a diagonal one $SO(5)_D$,  while  another  scalar field $\Phi_2 =\Omega_2 \phi_0$ located at the site-2 is responsible to break the $SO(5)_D$ into $SO(4)$.  Thus we obtain a total of $10+4$ NGBs.  In order to get rid of the  redundant  NGBs,   we are going to gauge a subgroup $ SU(2)_L \times U(1)_Y$ at the site-1 and the full  $SO(5)_2$ symmetry at the site-2.  Therefore  after the spontaneous symmetry breaking,  this two-site model delivers only  one copy of Higgs boson.

The two-site model  interprets the holographic nature of a composite Higgs  arising from a strong dynamics. We would like to illustrate this point by connecting  the deconstruction method with the CCWZ formalism.  Since  elementary fields in the SM  will be  put  at the site-1,  they play the roles of the source fields for operators constructed by  composite fields residing at the  site-2.  Let us simply set  $\Omega_2 =1 $ and  rescale  the kinetic terms to be  canonically normalized,  thus we obtain an effective  Lagrangian equivalent to the CCWZ description.  However  for the phenomenology relevance in the following discussion, we prefer  to  investigate the particle  spectrum in the unitary gauge, where only the  physical degrees of freedom appear in the Lagrangian.

\subsection{gauge sector}
First of  all, we  need to figure out the gauge interactions with those NGBs. In the unitary gauge, the pion fields in the coset of $SO(5)/SO(4)$ are parametrized in the sigma fields $\Omega_1$ and $\Phi_2$.  As pointed in the Ref.~\cite{Carena:2014ria},  there should be another scalar  field  $\Omega_X$, whose existence is necessary  to lift one combination of  extra $U(1)$ gauge fields.  Under the nearest  gauge interaction principle,  the  Lagrangian for the gauge bosons and nonlinear sigma fields is:
\beq
 \mathcal{L}_{2 - site} &=& \frac{f_1^2}{4}Tr{\left| D_\mu \Omega _1 \right|^2} + \frac{f_2^2}{2}{\left({D_\mu }{\Phi _2} \right)^T}{D_\mu }{\Phi _2}  +\frac{f_X^2}{4} \left| D_\mu \Omega _X \right|^2 \nonumber \\ &-& \frac{1}{4} Tr w_{\mu \nu} w^{\mu \nu} -\frac{1}{4} b_{\mu \nu} b^{\mu \nu} - \frac{1}{4}Tr{\rho _{\mu \nu }}{\rho ^{\mu \nu }} - \frac{1}{4} X_{\mu \nu} X_{\mu \nu} \,.
\eeq
with  the  covariant  derivative terms defined as:
\beq
D_\mu  \Omega_1 &=&  \partial_\mu \Omega_1   - i  g_0  w_{\mu}^a T^{aL} \Omega_1 -i g_0^\prime b_{\mu} T^{3R} \Omega_1 +  i g_\rho  \Omega_1  \rho_\mu^A T^A   \nonumber  \\
D_\mu \Omega_X &=& \partial_\mu \Omega_X -i g_0^\prime  b_\mu \Omega_X + i g_{X}   X_\mu \Omega_X \nonumber \\
D_\mu \Phi_2  &=&   \partial_\mu \Phi_2  - i g_\rho  \rho_\mu^A T^A \Phi_2
\eeq
where  $T^A $ is the generator for $SO(5)$ group and  $T^{aL}, T^{3R}$ are the generators in the subgroup $SU(2)_L$ and $SU(2)_R$.  The broken generator in the coset of  $SO(5)/SO(4)$ is denoted as $T^{\hat a}$, with $\hat a = 1, 2, 3, 4$. In the unitary gauge,  we  will simply set $\Omega_X =1$, thus the remaining  sigma  fields are parametrized as:
\beq
{\Omega _1} &=& \exp\left [ i  \frac{1}{f}  \frac{f_2^2}{f_1^2+ f_2^2}  \pi^{\hat a} T^{\hat a} \right],  \nonumber \\
{\Phi_2 } &=& \exp\left[ i  \frac{1}{f}  \frac{f_1^2}{f_1^2+ f_2^2}  \pi^{\hat a} T^{\hat a} \right] \phi_0,  \label{ome}
\eeq
with $\phi_0^t = (0,0,0,0,1)$ indicating a spontaneous symmetry breaking. Defining the field $\Phi= \Omega_1 \Phi_2$,  we obtain the physical sigma field  for this model set-up:
\beq
\Phi^T = \frac{1}{\pi} \sin(\pi/f_\pi)  \left ( \pi_1 ,  \pi_2, \pi_3, \pi_4, \pi \cot(\pi /f_\pi)  \right)
\eeq
Furthermore the following two-derivative  kinetic term is possible to add into the Lagrangian~\cite{DeCurtis:2011yx}, which is a non-local term,  but  allowed by the symmetries.
\beq
\mathcal{L}_{nl}  &=&  \frac{f_0^2}{2} \left( {D_\mu }{\Phi } \right)^T {D_\mu }{\Phi }  \nonumber \\
D_\mu \Phi &=& \partial_\mu \Phi - i  g_0 w_\mu^a T^{aL} \Phi - i g_0^\prime  b_\mu T^{3R} \Phi
\eeq
Since the non-local term contributes to  the pion kinetic term, it will modify  the pion decay constant.  Combining the results from $\mathcal{L}_{2-site} + \mathcal{L}_{nl} $ and demanding it  normalized according to $\frac{1}{2} \left( \partial_\mu \pi^{\hat a} \right)^2$,  we  obtain the  following  expression for  $f$:
\beq
f^2 = f_0^2 + \frac{f_1^2 f_2^2}{f_1^2+ f_2^2}\,.  \label{ef}
\eeq

The particle spectrum  for the gauge bosons  is  easy to be identified before the EWSB,  which are mildly corrected after setting  $\left\langle h \right\rangle \neq 0$. For  simplicity, we assume that $f_X = f_1$ and $g_X = g_\rho $,  the  mass eigenstates for those partial composite massive states are:
\beq
\tilde{W}^{\pm, 3}_\mu &=& \frac{1}{\sqrt{g_0^2+ g_{\rho}^2}} \left(g_0 w^{\pm, 3}_{\mu} - g_\rho \rho_{L\mu}^{\pm, 3} \right)\,,   ~~ B_{1\mu } = \frac{1}{\sqrt 2 }\left( \rho _{R\mu }^3 - {X_\mu } \right)   \nonumber \\
B_{2\mu } &=& \frac{1}{\sqrt {4g_0^{\prime 2} + 2g_\rho ^2} }\left( {g_\rho }\rho _{R\mu }^3 + {g_\rho }{X_\mu } - 2{g_0^\prime} b_\mu\right)
\eeq
with their masses squared calculated to be: $m_{\rho L}^2 = \frac{1}{2} (g_\rho^2 + g_0^2) f_1^2$, $m_{B1}^2 = \frac{1}{2} g_\rho^2 f_1^2$ and  $m_{B2}^2 = \frac{1}{2} (g_\rho^2+ 2 g_0^{\prime 2}) f_1^2$. There are another six massive  gauge bosons,  which do not mix with other fields at the leading order approximation.   The mass squared for  two charged ones $\rho_{R \mu}^{\pm} $ is  $m_{\rho R }^2 = \frac{1}{2} g_\rho^2 f_1^2$, whereas the mass squared for four axial ones $a_\mu^{\hat i } \,, \hat i = 1,2 ,3 ,4 $ is $m_a^2 = \frac{1}{2} g_\rho^2 (f_1^2 +f_2^2) $.  It should be noticed that with the existence of the non-local term,  we can set $f_2^2 < 0$,  and demanding  $f_1^2+ f_2^2> 0$ to ensure there are no tachyon modes,  which will lead to the condition $ f^2 < f_0^2$ as indicated by Eq.~[\ref{ef}].

\begin{figure}[!t]
\begin{center}
\includegraphics[angle=0,clip,width=7.0cm]{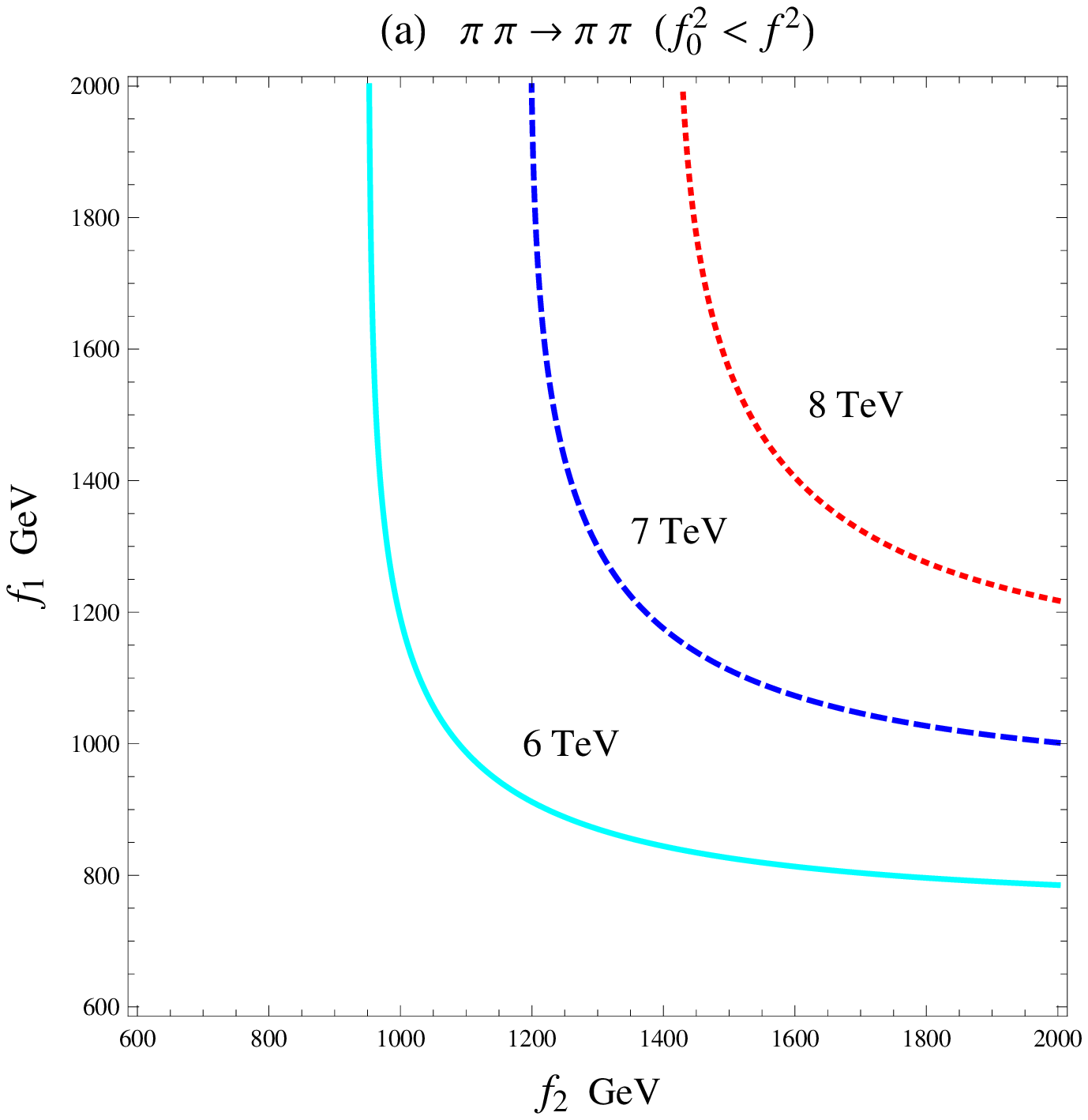}
\includegraphics[angle=0,clip,width=7.0cm]{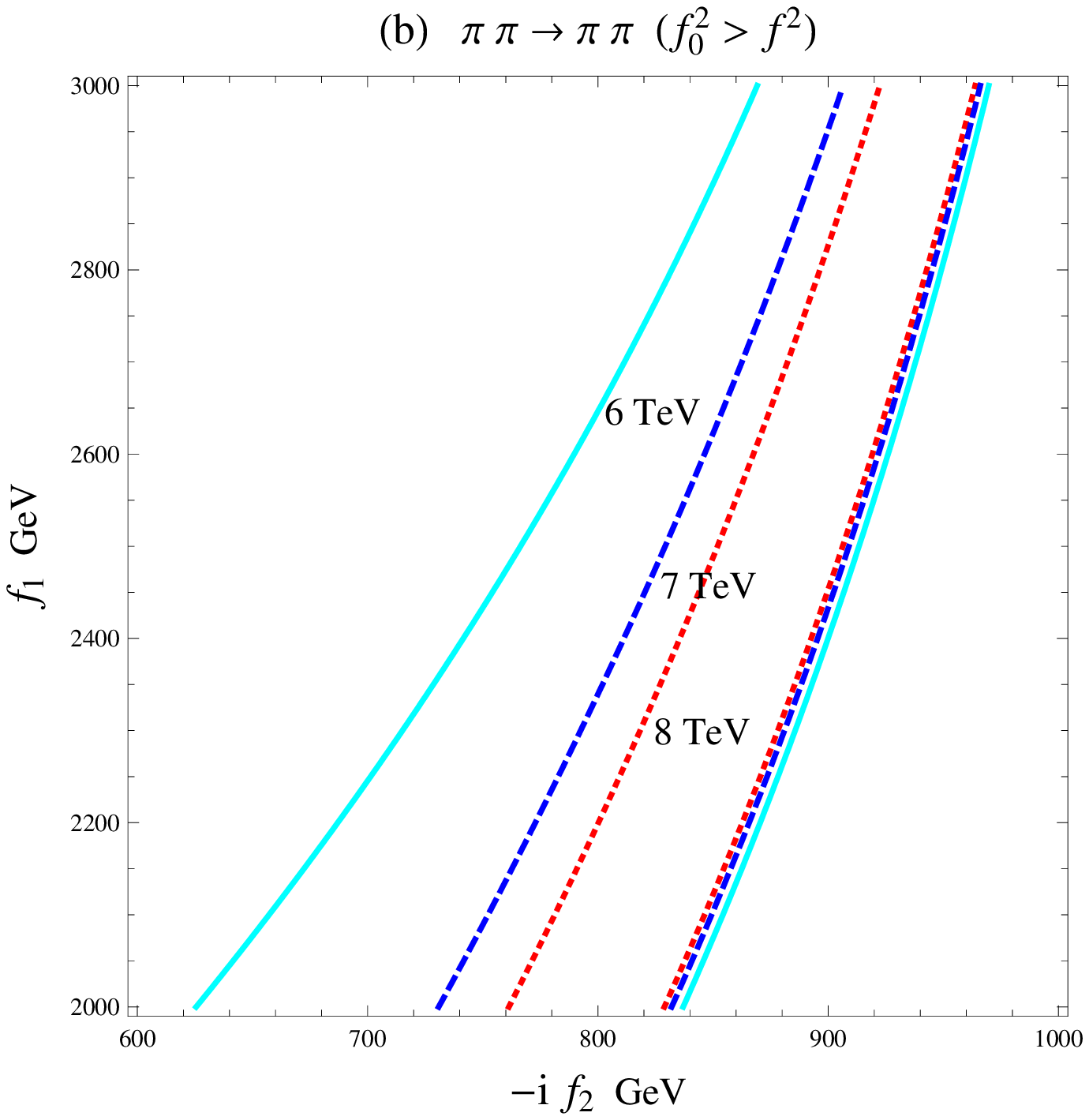}
\caption{{\small Parameter-space region where the unitarity bound
$|$Re$a^0_0(s)^{(\pi\pi)}|\leq \frac{1}{2}$ is violated at
energies $s\leq \Lambda^2$, for $\Lambda = 6.0\,, \,  7.0\,, \, 8.0$~ TeV, in the cyan, blue and red lines.  we fix the strong coupling to be $g_\rho =2.0$. The left panel is for  the case  with $f_2^2>0$ and $f_0 = 800~ \mbox{GeV}$, where the region  in the right and upper direction is allowed.  The right panel  is for the  case with $f_2^2< 0$ and  $f_0 = 1200~\mbox{GeV}$, where  the narrow region between the same color lines is allowed.}} \label{unitary}
\end{center}
\end{figure}

For the gauge  sector,  it is worthwhile to investigate  whether the unitarity for the pion scattering is partially restored in a two-site framework after the adding of  vector resonances.  Some  works in this direction have already been well done in~\cite{Contino:2011np, Bellazzini:2012tv, Cai:2013ira}.  We are interested to derive the $\rho_{L,R}$-$\pi$-$\pi$ and $\pi^4$ vertices which are relevant to the $\pi \pi \to \pi \pi$ scattering.  From the first term in $\mathcal{L}_{2-site}$, we can extract  the  interaction:
\beq
\mathcal{L}_{\rho \pi^2+\pi^4}^{(1)} & = &   \frac{(f^2-f_0^2)^2}{4 f^2 f_1^2} g_\rho\left[ {{\varepsilon ^{ijk}}{\pi ^i}{\partial _\mu }{\pi ^j}\rho _{L\mu }^k + \left( {{\pi ^k}{\partial _\mu }{\pi ^4} - {\pi ^4}{\partial _\mu }{\pi ^k}} \right)\rho _{L\mu }^k} \right]  \nonumber   \\
&+&   \frac{(f^2-f_0^2)^2}{4 f^2 f_1^2} g_\rho\left[ {{\varepsilon ^{ijk}}{\pi ^i}{\partial _\mu }{\pi ^j}\rho _{R\mu }^k - \left( {{\pi ^k}{\partial _\mu }{\pi ^4} - {\pi ^4}{\partial _\mu }{\pi ^k}} \right)\rho _{R\mu }^k} \right] \nonumber \\
&+& \frac{(f^2-f_0^2)^4}{24f_1^6 f^4}\left[ {{{\left( {{\pi ^a}{\partial _\mu }{\pi ^a}} \right)}^2} - {{\left( {{\pi ^a}{\partial _\mu }{\pi ^b}} \right)}^2}} \right]
\eeq
and from the second term in $\mathcal{L}_{2-site}$, we obtain a similar result:
\beq
\mathcal{L}_{\rho \pi^2+\pi^4}^{(2)} & = &  \frac{(f^2-f_0^2)^2}{2 f^2 f_2^2} g_\rho \left[ {{\varepsilon ^{ijk}}{\pi ^i}{\partial _\mu }{\pi ^j}\rho _{L\mu }^k + \left( {{\pi ^k}{\partial _\mu }{\pi ^4} - {\pi ^4}{\partial _\mu }{\pi ^k}} \right)\rho _{L\mu }^k} \right]    \nonumber \\
 &+&  \frac{(f^2-f_0^2)^2}{2 f^2 f_2^2} g_\rho \left[ {{\varepsilon ^{ijk}}{\pi ^i}{\partial _\mu }{\pi ^j}\rho _{R\mu }^k - \left( {{\pi ^k}{\partial _\mu }{\pi ^4} - {\pi ^4}{\partial _\mu }{\pi ^k}} \right)\rho _{R\mu }^k} \right] \nonumber \\
 & + & \frac{(f^2-f_0^2)^4}{6f_2^6 f^4}\left[ {{{\left( {{\pi ^a}{\partial _\mu }{\pi ^a}} \right)}^2} - {{\left( {{\pi ^a}{\partial _\mu }{\pi ^b}} \right)}^2}} \right]
\eeq
with the index  $i= 1, 2, 3$ and the indexes $a, b =1, 2, 3, 4$. While in the non-local term,  there exits additional $\pi^4$ self interaction term:
\beq
 \mathcal{L}_{\pi^4}^{nl} =  \frac{f_0^2}{6 f^4}\left[ {{{\left( {{\pi ^a}{\partial _\mu }{\pi ^a}} \right)}^2} - {{\left( {{\pi ^a}{\partial _\mu }{\pi ^b}} \right)}^2}} \right]
\eeq
Following the standard  procedure  described in~\cite{Cai:2013ira}, we get the partial wave expansion for the pion elastic scattering:
\beq
&& a^0_{0}(s)^{(\pi\pi)}
 = \frac{1}{16 \pi} \left( \frac{\left(f^2-f_0^2 \right)^4 }{4 f^4 f_1^6} +\frac{\left(f^2-f_0^2 \right)^4}{f^4 f_2^6} + \frac{f_0^2}{f^4}\right)  s  \nonumber \\
&& \quad +  ~\frac{g_s^2}{16 \pi } \left( \frac{\left(f^2-f_0^2 \right)^2}{2 f^2f_1^2} + \frac{\left(f^2-f_0^2 \right)^2}{f^2f_2^2} \right)^2 \bigg[\, \left(\frac{m_{\rho}^2}{s}+2 \right) \log \left(\frac{s}{m_{\rho }^2}+1\right)-1 \,\bigg]\, ,
\eeq
with the approximation $m_\rho = m_{\rho R} \simeq m_{\rho L}$ in the limit $g_0 \ll g_\rho$. Here we ignore the width effect from the vector resonance. Depending on the sign choice for the $f_2^2$,  two distinct scenarios  will occur for the unitarity  bound.   It is noticed  from the above equation that  when we choose  $f_2^2>0$,  the linear and logarithmic divergent terms  are both positive, which leads to the result that unless  $f_1$ and $f_2$ are large enough, the unitarity bound $|$Re$a^0_0(s)^{(\pi\pi)}|\leq \frac{1}{2}$ will be saturated  very quickly before the  effective cut  off scale is approached.   However in the other scenario $f_2^2 <0 $,  one should set  the  linear divergence  to be  almost vanishing, so that the high-energy behavior for the pion scattering is mainly determined by the mild logarithmic growing term.  In Fig.~\ref{unitary}, we plot the unitarity bound in the parameter space $(f_1, f_2)$ for  the two opposite situations by fixing the cut off scale to be a few TeV.   It turns out that  in  the case with $f_2^2>0$, adding a non-local term is kind of a benefit as it intends to enhance the ratio $f_1^2/ f_2^2$ (prefered by the $S$ bound) without too much raising the composite scale $f$.  On the other hand,  in the  case with $f_2^2<0$,   the partial cancelation in the linear $s$ term would help restore the perturbative unitarity.  But as we should observe from the figure,  it largely reduces the  unitarity conserving region as compared to the previous case. 
\subsection{fermion sector}
In this sector, we discuss  the embedding  of  fermions in the framework of a  two-site model.  Respecting  the full global symmetry,  the fermion is supposed to be put into  an  irreducible $SO(5)$ representation. Here  we are going to focus on the top quark sector   and  the simplest choice would be the basic representation. In order to let the fermions  acquire the right hypercharge assignments,  an extra  $U(1)_X$ symmetry is necessary such that    quantum numbers are determined by:  $Y = T_{3 R} + X$, and  $Q= T_{3 L} +T_{3 R} +X$. Since $SO(5)$ is spontaneously broken into $SO(4) \sim SU(2)_L \times SU(2)_R$,   the following decomposition  will apply: $5 =  (2,2) \oplus (1,1)$, which gives us one  bidoublet $\psi_{\pm \pm}$ with $T_{3L}, T_{3R}$ charges $(\pm \frac{1}{2}, \pm \frac{1}{2})$ and one singlet $\psi_{00}$. For the elementary fermions in the site-1, i.e. $q_L = (t_L, b_L)$, $t_R$ and $b_R$,  they  are embedded into incomplete $SO(5)$ representations with the non-dynamic spurion fields being turned off :
\beq
{\xi_L^u} = \frac{1}{{\sqrt 2 }}{\left( {\begin{array}{c}
{{b_L}}\\
{ - i{b_L}}\\
{{t_L}}\\
{i{t_L}}\\
0
\end{array}} \right)_{2/3}}\,, \qquad {\xi _R^u} = \left( {\begin{array}{c}
0\\
0\\
0\\
0\\
{i{t_R}}
\end{array}} \right)_{2/3}
\eeq
Let us simply introduce  one set of  composite fermions in the site-2, which should be accommodated in a complete $SO(5)$ representation:
\beq
\psi = \frac{1}{\sqrt 2 }\left( {\begin{array}{c}
X_4 +b_4   \\
i X_4 - i b_4 \\
- T_4 + t_4  \\
i T_4 + i t_4 \\ \sqrt 2 \, i T_1
\end{array}} \right)_{2/3}
\eeq
In the above construction we get one doublet  $(t_4, b_4)$, one non-standard doublet $(X_4,T_4)$, where the exotic quark $X_4$ carries an electric charge of $5/3$, and one singlet top $T_1$.

Due to the composite nature of  our Higgs field,  we can not directly  couple two SM fermions  with one Higgs field.   However a bilinear mix interaction is permitted, that is we  can use the link field $\Omega_1$ to connect the SM field in site-1 with the composite field in site-2. It is also possible for us to write down the $SO(5)$ invariant terms constructed with only the composite fermion fields and pNGBs, so that the SM  fermions will gain mass via the partial compositeness. As far as the top quark is concerned,  the Lagrangian for the fermion sector is:
\beq
\mathcal{L}_{top} &=& \bar q_L i {\not D}  q_L + t_R i {\not D} t_R  +\bar{\psi} i\not D^\rho  \psi  \nonumber  \\
&+& c _{tR} \bar \xi_R^u \Omega_1  {\psi_L} + c _{qL} \bar \xi_L^u \Omega_1 {\psi_R} - y_T {\bar \psi}_L \Phi_2 \Phi_2^T \psi_R - m_Y {\bar \psi}_L {\psi}_R+ h.c.  \label{top}
\eeq
where after the EWSB in the unitary gauge, the $\Omega_1$ takes the following  simple form:
\beq
\Omega_1 = \left( \begin{array}{ccc}
I_{3 \times 3} & & \\
&  \cos \frac{f}{f_1^2} h & \sin \frac{f}{f_1^2} h  \\
&  -\sin \frac{f}{f_1^2} h & \cos \frac{f}{f_1^2} h
\end{array} \right)\, ,
\eeq
and the explicit expression for the other scalar field is:
\beq
\Phi_2^t =\left( 0,0,0,\sin \left[\frac{f_1^2}{f_1^2+f_2^2} \frac{h}{f}\right],\cos \left[\frac{f_1^2}{f_1^2+f_2^2} \frac{h}{f} \right]\right) \,.
\eeq
There are four top quarks with electric charge $2/3$ in this two-site model: $(t, t_4, T_4, T_1)$, with the relevant mass term:
\beq
\mathcal{L}_m = \overline {\left( {\begin{array}{c}
t_L\\
t_L^4\\
T_L^4\\
T_L^1
\end{array}} \right)}  M_{top}
\left( {\begin{array}{c}
t_R\\
t_R^4\\
T_R^4\\
T_R^1
\end{array}} \right) + h.c.
\eeq
Since  $v < f_{1,2}, f$,  let us expand  the $M_{top}$ to the order of $\mathcal{O} (v/f)$,
\beq
M_{top} =\left(
\begin{array}{cccc}
 0 & c_{q L} & 0  & \frac{c_{q L} f_2^2 v}{\sqrt{2} f (f_1^2+f_2^2)} \\
 -\frac{c_{t R} f_2^2 v}{\sqrt 2 f (f_1^2+f_2^2)}  & -m_Y & 0 & -\frac{y_T f_1^2 v}{\sqrt{2} f (f_1^2+ f_2^2)} \\
 -\frac{c_{t R} f_2^2 v}{\sqrt 2 f (f_1^2+f_2^2)}  & 0 & -m_Y & -\frac{y_T f_1^2 v}{\sqrt{2} f (f_1^2 + f_2^2)} \\
 c_{t R} &  -\frac{y_T f_1^2 v}{\sqrt{2} f (f_1^2 + f_2^2)}  &  -\frac{y_T f_1^2 v}{\sqrt{2} f (f_1^2 + f_2^2)} &  -m_Y-y_T
\end{array}
\right) \,.
\eeq
The top quark mass matrix is easy to be analytically diagonalized if we set the Higgs VEV to be zero. From the Lagrangian $\mathcal{L}_{top}$, the mass matrix for the two bottom quarks $(b, b_4)$  can also be extracted.  But unlike the top quark case,  the bottom quark mass matrix has no dependence on the Higgs field.   Since the mixing pattern for the left  handed  top and  bottom quarks  coincides with each other at $\left\langle h \right\rangle = 0$,   the following rotation simultaneoulsy tranforms them into the mass eigenstates:
\beq
{\tilde t}_L &=& \frac{m_Y}{\sqrt {c_{qL}^2 + m_Y^2} } t_L + \frac{c_{qL}}{\sqrt {c_{qL}^2 + m_Y^2} }  t_{4L} \,, \nonumber \\
{\tilde t}_{4 L} &=& -\frac{c_{qL}}{\sqrt {c_{qL}^2 + m_Y^2} } t_L + \frac{m_Y}{\sqrt {c_{qL}^2 + m_Y^2} } t_{4L} \,,
\\
{\tilde b}_L &=& \frac{m_Y}{\sqrt {c_{qL}^2 + m_Y^2} } b_L + \frac{c_{qL}}{\sqrt {c_{qL}^2 + m_Y^2} } b_{4L} \,, \nonumber  \\
{\tilde b}_{4 L} &=& -\frac{c_{qL}}{\sqrt {c_{qL}^2 + m_Y^2} } b_L + \frac{m_Y}{\sqrt {c_{qL}^2 + m_Y^2} } b_{4L} \,.
\eeq
While for the right handed top quarks,  it is the two singlets $(t_R, T_{1R})$ that would mix with each other and the corresponding rotation is:
\beq
{\tilde t}_R &=& \frac{m_Y+ y_T}{\sqrt {c_{tR}^2 + (m_Y+y_T)^2} } t_R + \frac{c_{tR}}{\sqrt {c_{tR}^2 + (m_Y+y_T)^2} } t_{1 R}  \,, \nonumber  \\
{\tilde t}_{1 R} &=& -\frac{c_{tR}}{\sqrt {c_{tR}^2 + (m_Y+y_T)^2} } t_R + \frac{m_Y+y_T}{\sqrt {c_{tR}^2 + (m_Y+y_T)^2} } t_{1 R}  \,.
\eeq
Defining the mixing angles $\sin \theta_L = \frac{c_{qL}}{ \sqrt{c_{qL}^2 + m_Y^2}}$ and  $\sin \theta_R = \frac{c_{tR}}{ \sqrt{c_{tR}^2 + (m_Y+ y_T)^2}}$,  thus to the leading order expansion,  the top quark mass can be  approximated as,
\beq
m_{t0} \simeq  \frac{|y_T| \sin \theta_L \sin \theta_R  }{\sqrt{2} } \frac{v}{f}  \,.
\eeq
where  $\sin \theta_L$ and $\sin \theta_R$  indicate  ￼￼￼the degree of compositeness for $t_L$ and $t_R$ respectively.  Notice that  though  the term proportional to $y_T$ in Eq.~[\ref{top}] only gives mass to a singlet top $T_1$,  it  is necessary to be present  for the SM top $t_0$ to gain  the observed mass. Furthermore,  suppose that $|y_T|$  is  of  a  few  TeV energy scale,  we need  the left  handed or the  right handed top to mostly origin from the composite sector. The  masses for three heavy top quarks are determined by:
\beq
m_{t4}^2 = c_{qL}^2 + m_Y^2 \, , ~~ m_{T4}^2 = m_Y^2 \,,  ~~ m_{T1}^2 = c_{tR}^2 + (m_Y + y_T)^2
\eeq
For the composite sector,  we generally will set  the parameter $y_T$ to be positive.  In such a case, the $SU(2)_L$ partner for $T_4$,  i.e. an exotic quark $X_4$, would be the lightest fermionic resonance, since it gets no further correction of  $\mathcal{O} (v/f)$ after the EWSB.  On the other hand,  when we choose a negative $y_T$,  the  lightest fermion could either be  the  singlet top $T_1$ or  the exotic  quark $X_4$.

\section{ Electroweak Constraint from $S$ and $T$}
Oblique parameters associated with the electroweak  precision  test  puts  a  severe bound on the parameter space  for ``universal" models beyond the Standard Model. Let us first recall the definitions of  oblique parameters, which are extracted from  the  two-point functions of  weak currents for  the gauge bosons.  $S$, $T$ and $U$ correspond to the residue coefficients for expansion up to the order of $p^2$ after fixing gauge couplings,  Higgs VEV  and imposing the $U(1)_{em}$ gauge invariance~\cite{peskin, Barbieri:2004qk}.  Roughly speaking,  heavy fields from the EW symmetry breaking sector additively contribute to the $S$,  while the effect of  isospin breaking is  counted by the $T$ and $U$. In the two-site composite Higgs model, there is a tree level mixing between the elementary gauge fields and composite gauge fields.  Thus after integrating out heavy spin-$1$ resonances (both the vector and axial bosons), we find the dominating contribution to  $S$ parameter is:
\beq
&& \Delta S = -16 \pi \cdot  {\Pi '}_{W^3B} (0) \nonumber \\ &=& \frac{8 \pi}{g_\rho^2} \left[ 1 - \frac{f_1^4}{\left( {f_1^2 + f_2^2} \right)^2}\right]\frac{v^2}{f^2}
\eeq
\begin{figure}[!t]
\begin{center}
\includegraphics[angle=0,clip,width=7.0cm]{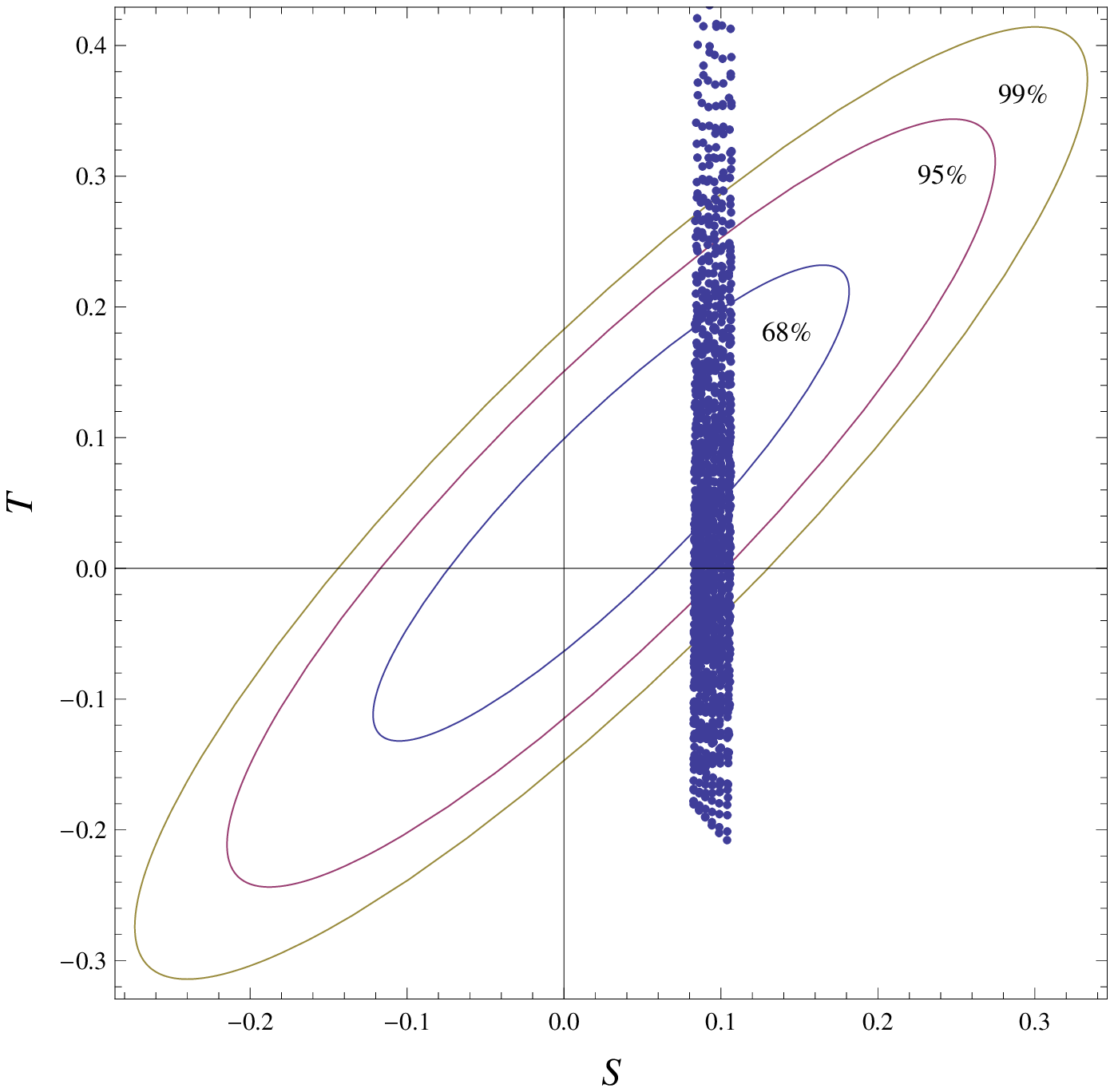}
\includegraphics[angle=0,clip,width=7.0cm]{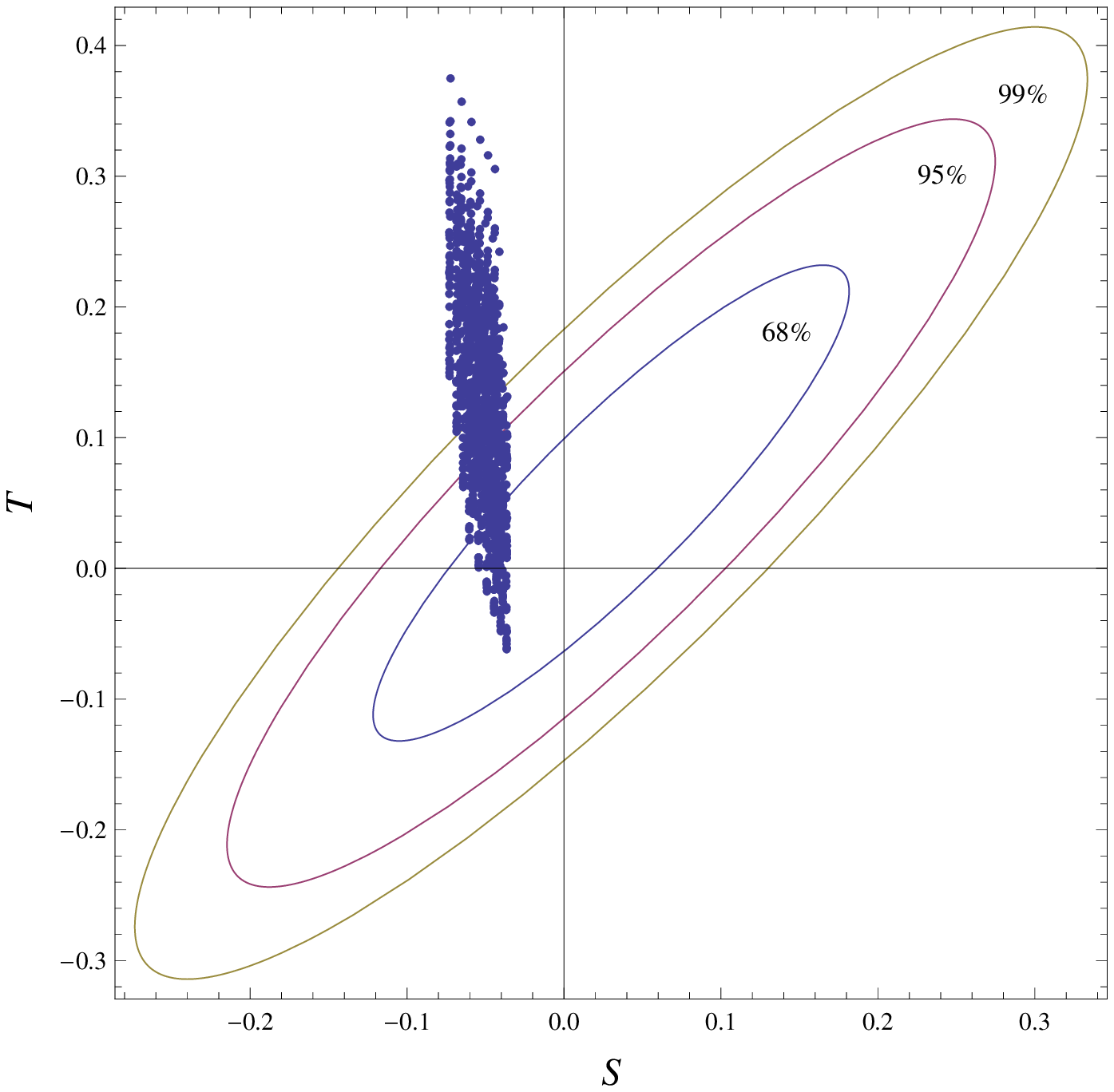}
\caption{{\small S-T plane for  the composite resonance parameter scanning. The ellipses are at the $68 \% ~(1 \sigma)$, $95\% ~(2 \sigma)$ and $99 \% ~(3 \sigma)$ confidence levels. The strong coupling is fixed to be $g_\rho =2.0$. In the left contour parameters are  in  the range (GeV): $3000< c_{tR}, y_{T}<3200$, ~$ 3000<m_{Y}, f_{1}<3500$, $ 800<f_{2} <830$ and $f_0 = 0.0$;  In  the right contour parameters are in the range (GeV): $3400< c_{tR}<3600$, ~$3000< y_{T}<3200$, ~$ 3000<m_{Y} , f_{1}<3500  $, $ 800< - i f_{2} <830$ and $f_0 = 1200$. The remaining parameter $c_{qL}$ is calculated  with an input  top quark mass and the points which are located in the ellipses pass  the  EWPT. }}
\end{center}
\label{ST}
\end{figure}
As we should notice  that  although this  effect is proportional to  $v^2/f^2$,  the tree level deviation still  imposes a  stringent constraint due to  a factor of  $8 \pi$ unless the strong coupling $1<g_\rho < 4\pi$ is  large enough.  It is argued in ref.~\cite{Agashe:2007mc} that the  $S$ parameter is  generally positive  in  most  extra dimension scenarios. However,  in  the two-site model with the existence of  non-local term,   it is possible  to enforce  the condition $0  <f_1^2 + f_2^2 < f_1^2 $, thus we obtain a negative  $S$ parameter. Moreover we  have the freedom to ensure $ |f_2| \ll |f_1|$,  therefore the leading order contribution to the $S$ parameter is further suppressed, with the magnitude of  its deviation being kept in the vicinity of  $0.1$.

Since those composite vector resonances remain in the irreducible $SO(4) \simeq SU(2)_L \times SU(2)_R$ representations after the spontaneous  global symmetry breaking $SO(5) \to SO(4)$, the tree level deviation to  $T$ and $U$  is  zero  due to  the custodial symmetry protection.  Nonetheless the correction will arise at the loop level.
\beq
\Delta T = \Delta U = 0
\eeq
Another calculable source for $S$ and $T$ origins from the reduced gauge couplings with Higgs boson,  which give rise to the infrared (IR) contribution. It is convenient to derive those couplings in the holographic basis by expanding the Higgs field around its VEV,
\beq
&& \frac{1}{4} g^2 f^2  W_\mu^+ W_\mu^- \sin ^2\left(h/f\right)  + \frac{1}{8}\left( g^2+ g^{\prime 2} \right) f^2 Z_\mu Z_\mu \sin ^2\left(h/f\right)  \nonumber \\
&\Rightarrow&  \frac{1}{2} \frac{2m_W^2}{v} ( v+ a ~ h ) W_\mu^+ W_\mu^-  + \frac{1}{2} \frac{m_Z^2}{v} (v+ a ~ h) Z_\mu Z_\mu  + \mathcal{O}(h^2) \, .
\eeq
with  the  masses of  gauge bosons and  the  parameter $ |a| <1$  determined by:
\beq
 m_W^2 = \frac{g^2 {f^2}\sin^2 (v/f)}{4}, ~ m_Z^2 = \frac{( g^2 + g^{\prime 2} ){f^2}\sin^2 (v/f)}{4},  ~ a = \cos (v/f) \eeq
In the SM, there is an exact cancelation of  logarithmic divergence for the $S$ and $T$, which is spoiled by the  reduced Higgs  couplings with gauge bosons.  Therefore  the  IR contribution  in fact  describes  a running effect  from the  EW scale till the composite scale in the  effective theory,  where  by the  NDA estimation $\Lambda_{NDA}= 4 \pi f$.
\beq
\Delta {S_{IR}} &=& \frac{1}{6 \pi} \left[ \sin ^2\left(v/f\right) \log \left( \frac{\Lambda }{m_h} \right) + \log \left( \frac{m_h}{m_{h,ref}}\right)\right]   \,  ,    \\
\Delta {T_{IR}} &=&  - \frac{3  }{8 \pi c_w^2} \left[ \sin ^2\left(v/f\right) \log \left( \frac{\Lambda }{m_h} \right)   + \log \left( \frac{m_h}{m_{h,ref}}\right) \right]
\label{IR} \eeq
Under the condition  that $f \sim  1.0$ TeV,  i.e. assuming  no large splitting exits between the Higgs VEV and the composite scale $f$,  the IR contribution generally constitutes a sizable portion to both $S$ and $T$.

Finally let us discuss more about the fermion loop correction to EW precision test.  The evaluation for  the  $S$ and $T$ from various  types of  vector-like quarks is given in the reference~\cite{Lavoura}. Some detail studies of  $T$ parameter constraint on vector-like singlet, doublet and triplet quarks could also be found in~\cite{Cacciapaglia:2010vn, Cai}. I  will give the  analytic expressions of  $S$ parameter from vector-like quark loops, especially for the nonstandard doublet scenario in the appendix.  It is worth to point out  that since the exotic quark $X_4$ carries an electric charge of  $\frac{5}{3}$,  the function $\psi_{+}$ defined in the original work~\cite{Lavoura} will be modified.  For the $S$ parameter,  the virtual fermion effect is generally  subleading, but not  totally irrelevant,  which  leads to the consequence that the $S$ is less model dependent  on  the mixing parameters.  By contrast the  fermion contribution  to the $T$ is more important and needs to take into consideration. We expect that the $T$ obtains a substantial positive shift  though quark mixings, so that  it  will  partially compensate  the negative  IR  correction. The  magnitude of  $T_{ferm}$ in a generic composite Higgs model can be estimated in the limit $|f_2| \ll |f_1|$ as:
\beq
\Delta T_{ferm} \sim \frac{N_c}{4 \pi s_W^2 g^2} \frac{ y_T^4 \sin^4 \theta_{L,R} }{M_T^2 f^2} \sin \frac{v^2}{f^2}\,,
\eeq
where $M_T$  collectively stands  for the mass of a composite vector-like quark. In order to get a quantitative understanding,  we could  assume that $\sin \theta_{L,R} \sim  0.6 $,  $M_T \sim |y_T| \sim 3.0$ TeV, and  $f \sim  1.0 $ TeV,   thus  an estimation is obtained $\Delta T_{ferm} \sim 0.17 $,  which is numerically competitive with the IR correction. The sign for the $T$ parameter is determined by the isospin of  the heavy top quark.  As we have shown in the mass matrix for top-like quarks,  the mixing  is entangled with each other.  For simplicity, we consider that the top quark separately mixes with one type of  vector-like quark each time.  Using the equations presented in~\cite{Cai},  it is possible to exactly evaluate the $T$ parameter for  various scenarios.  For the SM top $t_0$ mixing with  one  singlet  $T_1$,  or mixing with one doublet $(t_4, b_4)$,   $\Delta T_{ferm}$ is always positive.  The situation becomes  opposite for the nonstandard  doublet  $(X_4, T_4)$, as its modification to the $W_3^\mu W_3^\mu$ form factor is  bigger than the  other two cases. In the small mixing limit,   the last type mixing gives a negative contribution  to  the $T$ parameter.  In this two-site model,  we can find  that because the $t_0$ mixing with $T_4$ is  larger than with $t_4$, under the condition $M_{T4} < M_{t4}$,  the negative contribution from nonstandard doublet will overcome the positive one from the doublet. However provided that  additional  positive contribution obtained from  the mixing with $T_1$ is large enough,  the combining result  $\Delta T_{ferm}$  would  still be  positive, as we can prove this point using the numerical scanning.

In Fig.~\ref{ST},  we show  the numerical results for  $S$ and $T$ by  scanning the  parameter set $( c_{qL}, c_{tR}, m_Y, y_T, f_0, f_1, f_2)$  in typical ranges.  It is observed that  the $S$ is mainly sensitive to $f_0$ and $f_{1,2}$, whereas the $T$ is  more dependent on the mixing parameters for the fermions. We are intending to find out  viable parameter regions which are compatible with  experimental values, i.e. $ S =0.03 \pm 0.10$ and $ T= 0.05 \pm 0.12$  with a correlation coefficient of  $\rho_{col} = 0.89$.  In the left contour,  we illustrates one case with no non-local term, i.e. $f_0= 0$,  therefore the $S$ parameter is  bound to be positive due to the condition $f_1^2 < f_1^2 + f_2^2$.  Combining all the mixing effects from the SM chiral tops with composite vector-like quarks,  we find out that there is enough possibility for the $T$ to  be shifted into the positive region. While in the right contour,  another case  with a non-local term is present.  Since we enforce an opposite condition $f_1^2> f_1^2 + f_2^2$,  the sign of $S$ parameter is tuned to be negative.  In the latter case,  the  points with a negative $T$ are more consistent with  the electroweak data.

\section{Gluon fusion to Higgs Production}
Since the Yukawa couplings for the composite fermions are crucial for the gluon fusion process, we are going to briefly comment  their contribution to the Higgs production.  For all the fermion fields in  the mass eigenstates,  let us assume that they are interacting with the Higgs field in the following way:
\beq
\mathcal{L}_h = \sum {M_i} \bar \psi_i \psi_i + \sum {Y_{ii}} h^0 \bar \psi_i \psi_i
\eeq
\begin{figure}[!t]
\begin{center}
\includegraphics[angle=0,clip,width=7.5cm]{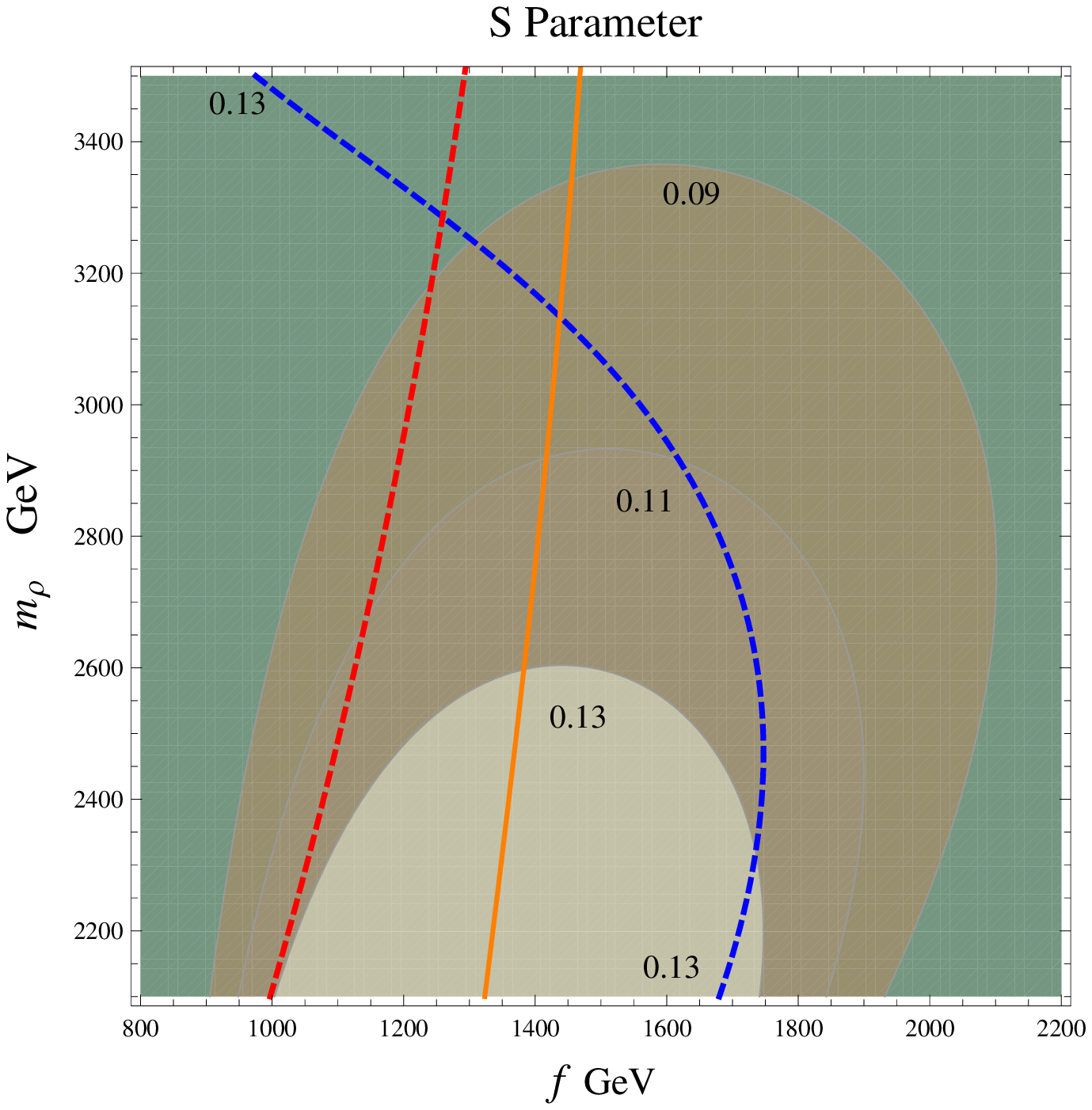}
\includegraphics[angle=0,clip,width=7.5cm]{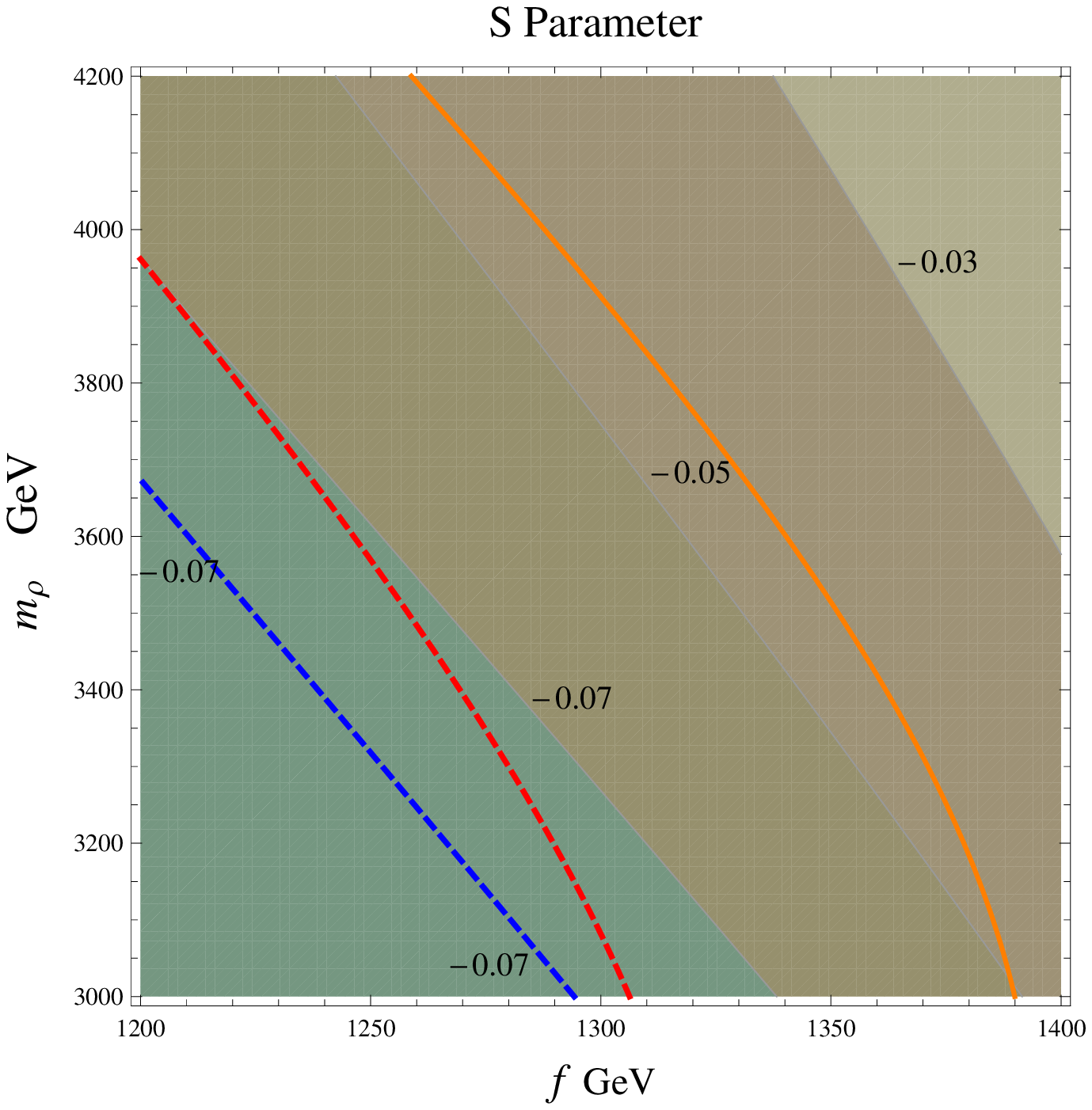}
\caption{{\small S parameter contour and unitarity bound with $\Lambda= 8 $ TeV on the plane of $(m_\rho, f)$. In the left contour, we set  $f_0 =800$ GeV  and  demand that  $f_1^2+ f_2^2 >f_1^2$.  The orange line corresponds to the unitarity bound, where  the region in the right direction is allowed.  The $S=+0.13$ bound (blue dashed line)  and the unitarity bound (red dashed line)  at  $f_0 =0.0$ GeV  is shown for comparision. While  in the right contour, we set $f_0 = 1600$ GeV and instead enforce $f_1^2 + f_2^2 < f_1^2$.  The orange line represents the unitarity bound, where the region in the lower direction is allowed.  The $S= -0.07$ bound (blue dashed line) and the unitarity bound (red dashed line) at $f_0= 1550$ GeV is  shown for comparision.}} \label{spic}
\end{center}
\end{figure}
where $M_i$ and $Y_{ii}$ are the mass and Yukawa coupling for  each fermion. Notice that in our simplified model,  only  the top-like quarks have interaction with the Higgs field.  The production rate for a  Higgs boson from  fermion loops is proportional to:
\beq
\sigma \left({gg \to h^0} \right) \propto \left| \sum\limits_i \frac{Y_{ii}}{M_i} A_{1/2}\left( \tau _i \right) \right|^2 \,, ~~  \tau_i = \frac{m_h^2}{4M_i^2}
\eeq
For the case of a light Higgs boson,  we generally have $M_{Q, t} \gg m_h \gg m_b$, with the index $Q$ stands for the heavy vector-like quarks.  In order to make contact with the low energy phenomenology,  one just needs to  take advantage of  the  approximation for $A_{1/2} \left (\tau_i \right)$ in certain limits, i.e. $A_{1/2} \left (\tau \right) \to 4/3$, for $ \tau \to 0$,  and $ A_{1/2} \left (\tau \right) \to 0 $, for $\tau \to \infty$.  Furthermore relating  the sum of   $ \sum y_{ii} / M_i$ to the  determinant of the mass matrix:
\beq
\sum\limits_i \frac{Y_{ii}}{M_i} = \frac{\partial \log \left( \det M \right)}{\partial v} \,, \eeq
we can effectively evaluate the  production rate without rotating into the mass eigenstate.  Since the $X_{5/3}$ and $b_4$ in this model do not couple to the Higgs field,  by neglecting the bottom quark contribution we obtain a concise result :
\beq
\frac{\sigma ( gg \to h^0 )_{CHM}}{\sigma (gg \to h^0)_{SM}} \simeq \frac{2 v}{f} \cot \left(\frac{2 v}{f}\right) = 1-\frac{4}{3} \frac{ v^2}{ f^2}
\eeq
Therefore in the case  with only  one multiplet of composite quarks,  the gluon fusion production of Higgs boson is  generally reduced with respect to the  SM  scenario,  as pointed out by the pioneer work~\cite{Falkowski:2007hz}.   However it is noticed in ref.~\cite{Azatov:2011qy}  that through introducing  the $h$ dependent bottom quark mixing with composite quarks,  an enhanced  $h^0 gg$ coupling is possible to be realized in the composite Higgs scenario.

Through constraining the decay constant using the unitarity and  EWPT,  we will be able to estimate the reduced percentage in this model.  The $S$ parameter,  especially its dominate part  $\Delta S_{fit} \simeq \Delta S_{tree} + \Delta S_{IR}$,  imposes a stringent bound on $m_\rho$ and $v^2/f^2$ by  requiring that  $ -0.07 < S < 0.13 $. On the other hand, we should also require the unitarity bound to be conserved till a  relative  large effective cut off scale, e.g. $8$ TeV.  The Fig.~\ref{spic} interprets the $S$ parameter  and unitarity bound in the $(m_\rho, f)$ plane.   For the case $f_1^2+f_2^2>f_1^2$,  it turns out that  a larger $f_0$  leads to  a growing  $f$,  but  the mass scale $m_\rho$  will be lowered, resulting in  more  parameter region compatible with the experimental data.  However, for the opposite case $f_1^2+ f_2^2 <f_1^2$,  the unitarity  bound intends to  push  $f$  to be smaller, while the  $S$  parameter  requires a larger $f$.  Thus one has to increase the $f_0$ in order to gain more  compatible region.   Under either  situation,  we  will find out that  after imposing necessary  constraints,  an upper bound  $v^2/f^2 < 0.042$ is  generally permitted, which in turn translates into a  rough  estimation  $\sigma_{CHM}/ \sigma_{SM} > 0.94$ for the reduced Higgs production rate.  Notice that  a lower production rate is applicable  if one  reduces  the $\Lambda_{eff}$  for the unitarity bound.  

\section{Conclusion}
In this paper, we study the minimal $SO(5)/SO(4)$ Composite Higgs Model  in a two-site scenario, where only the first level of  composite resonances is present in comparision with the KK modes in an extra dimension theory.  In addition to the nearest neighbor interaction,  we especially investigate the effects of  non-local term on the  perturbative unitarity  and  the EW precision test.  In the scenario $f_2^2>0$,  the existence of  non-local term will lead to a  lower bound for the vector rosonance mass $m_\rho$, which  ameliorates the  compatibility of this two-site model with  experimental data.   On the other hand,  the non-local term is necessary to be added  in  the scenario $f_2^2<0$.  Under such a situation,  there  is  a tension between the unitarity  requirement and the negative $S$ parameter  bound.  Thus  we should  increase the $f_0$  in order to relieve this tension and achieve a  relative large  effective cut off scale $\Lambda_{eff}$.  

For simplicity,   vector-like composite fermions are embedded in  the  $SO(5)$ basic representations, which will mix the SM quarks via  bilinear  interactions.  Under the situation of  a positive  $S$ parameter,  we  prefer  a positive $T$  for a better fit with the EW precision test. Through the  parameter scanning,  we have shown that there is enough possibility that the positive correction to the $T$  from vector-like fermions could dominate the  negative IR  contribution from the reduced  Higgs couplings.  Furthermore, the  effects of  vector-like fermions on the  gluon fusion Higgs production is  discussed.  It turns out the  production rate will at most be  reduced around  $6 \%$  provided that we demand a  strict  unitarity and  EWPT to be satisfied.

\section*{Acknowledgments} 
H.Cai is  supported by the postdoctoral foundation
under the Grant No. 2012M510001, and  in part supported by National Nature Science
Foundations of China (NSFC) under the Contract No. 10925522.

\newpage

\section*{Appendix}
In this appendix,  we are going to collect the contribution to $S$ parameter from vector-like fermions in  singlet,  doublet and nonstandard  doublet scenarios. Consider the top quark mixing  with each type of  vector-like  fermion in the following way:
\beq
\mathcal{L}_{top}  &\supset& - m_t \bar t_L  t_R  -  x_{T_1} \bar t_L T_{1R} - x_{t4} \bar  t_R  t_{4 L}  - x_{T4} \bar  t_R T_{4 L} \nonumber \\ & - & M_{T_1} \bar T_1  T_1- M_{t_4} \bar t_4  t_4 - M_{T_4}  \bar T_4 T_4 + h.c.
\eeq
For a singlet vector-like fermion,  the  mixing angles are determined by diagonalizing the mass matrix,
\beq
\sin {\theta_{u}^L} = \frac{M_{T_1} x_{T_1}}{\sqrt {(M_{T_1}^2 - m_t^2)^2+ M_{T_1}^2x_{T_1}^2}} \,, ~~~
\sin {\theta _{u}^R} = \frac{x_{T_1}}{M_{T_1}} \sin  {\theta_{u}^L} .
\eeq
While  for  a doublet or  nonstandard doublet,  the L.H. and R.H. mixing angles are  exchanged with respect to the singlet  scenario,
\beq
\sin {\theta _{u}^R} = \frac{M_{t_4(T_4)} x_{t_4(T_4)}}{\sqrt {(M_{t_4(T_4)}^2 - m_t^2)^2} + M_{t_4(T_4)}^2 x_{t_4(T_4)}^2}\,, ~~~\sin {\theta _{u}^L} = \frac{x_{t_4(T_4)}}{M_{t_4(T_4)}} \sin  {\theta_{u}^R} .
\eeq
In terms of mixing angles, the $S$ parameter  for each kind of scenario can be expressed by the following equations:
\beq
\Delta S_{t_0-T_1}  &=&  \frac{3}{2 \pi } \left [ \sin^2 \theta_u^{L  }{\psi_ + }({y_{T1}},{y_b}) - \sin^2 \theta_u^{L  } {\psi_ + }({y_t},{y_b}) \nonumber \right.  \\ &-& \left.   \cos^2 \theta_u^{L  } \sin^2 \theta_u^{L  } {\chi_ + }({y_{T1}},{y_t})  \right]  \nonumber  \\
 \Delta S_{t_0-t_4}  &= &  \frac{3}{2 \pi } \left[ \sin^2 \theta_u^L {\psi_ + }(y_{t4},y_b) -  \sin^2 \theta_u^L {\psi_ + }(y_t , y_b)  \nonumber \right. \\ &+ & \left.
(  \cos^2 \theta_u^L+ \cos^2 \theta_u^{R } ) {\psi_ + }(y_{t4} , y_{b4})   +
 (\sin^2 \theta_u^L + \sin^2 \theta_u^{R } ) {\psi_ + }(y_t,y_{b4})    \nonumber \right.  \\ &+ &
 \left.      2   \cos \theta_u^L  \cos \theta_u^{R } {\psi_ - }(y_{t4} , y_{b4})   +   2  \sin \theta_u^L \sin \theta_u^{R }   {\psi_ - }(y_t , y_{b4})   \nonumber \right. \\ &-&  \left. \cos^2  \theta_u^{R } \sin^2  \theta_u^{R } {\chi_ + }(y_t,y_{t4}) \right]    \nonumber  \\
\Delta S_{t_0-T_4}  &= & \frac{3}{2 \pi } \left[ \sin^2  \theta_u^{L }{\psi_ + }(y_{T4},y_b)  - \sin^2 \theta_u^{L  } {\psi_ + }(y_t, y_b)
  \nonumber  \right. \\  &+ &  \left.   ( \sin^2  \theta_u^{L } + \sin^2  \theta_u^{R }  ) {\psi_ + }(y_{X4}, y_t ) +  (\cos^2  \theta_u^{L } + \cos^2  \theta_u^{R }  ) {\psi_ + }(y_{X4}, y_{T4} )  \nonumber \right.  \\ &+ &
 \left.  2  \sin \theta_u^{L } \sin \theta_u^{R }  {\psi _ - }(y_{X4}, y_t ) +   2   \cos \theta_u^{L } \cos \theta_u^{R }   {\psi_ - }( y_{X4}, y_{T4} )   \nonumber \right. \\ &-&
\left.( 4 \cos^2 \theta_u^L \sin^2 \theta_u^L  + \cos^2 \theta_u^{R } \sin^2 \theta_u^{R } ) {\chi_ + }(y_t,y_{T4})  \nonumber \right. \\ &-&
\left. 4 \cos \theta_u^L \sin \theta_u^L   \cos \theta_u^{R } \sin \theta_u^{R }  {\chi_ - }(y_t,y_{T4}) \right]    \label{spara}
\eeq
where the rescaled mass squared is defined as  $y_i = M_i^2 / m_Z^2$,  with $M_i$ representing the mass of  vector-like quark and the functions of $\chi_+ $, $\chi_-$, $\psi_+$ and $\psi_-$ are:
\beq
\chi _+ \left(y_1, y_2\right) &=& \frac{5\left(y_1^2 + y_2^2\right) - 22 y_1 y_2}{9\left( y_1 - y_2 \right)^2} + \frac{3y_1 y_2 \left( y_1 + y_2 \right) - y_1^3 - y_2^3}{3\left(y_1 - y_2 \right)^3}\ln \frac{y_1}{y_2}\,,  \nonumber \\
\chi_- \left(y_1, y_2\right) &=&  - \sqrt {y_1 y_2} \left( \frac{y_1 + y_2}{6y_1 y_2} - \frac{y_1 + y_2}{\left( y_1 - y_2\right)^2} + \frac{2 y_1 y_2}{\left( y_1 - y_2 \right)^3} \ln\frac{y_1}{y_2} \right)\,,  \nonumber \\
\psi _+ \left(y_\alpha, y_i \right) &=& \frac{1}{3} - \frac{1}{3}\left(Q_\alpha + Q_i\right)\ln \frac{y_\alpha}{y_i} \,,  \quad
\psi _- \left(y_\alpha, y_i \right) =  - \frac{y_\alpha + y_i}{6\sqrt {y_\alpha y_i} } \,.
\eeq
where our function $\psi_{+} (y_\alpha, y_i)$, with the index $\alpha$ for an up-type quark and the index $i$ for a down-type quark,  is in fact dependent on the sum of electric charges $Q_\alpha+Q_i$,  which  generalizes the previous result reported  in  the Ref.~\cite{Lavoura}. Notice that for the mixing of  $t_0$ with $T_4$,  the argument $y_{X4}$ should be put in front of the argument $y_{T4}$ or $y_t$, due to the opposite  isospin  assignment in a non-standard doublet.

\end{document}